# Time-on-Task Estimation with Log-Normal Mixture Model

**Ilia Rushkin**
Harvard University
Cambridge, MA, USA
ilia_rushkin@harvard.edu

## ABSTRACT
We describe a method of estimating a user's time-on-task in an online learning environment. The method is agnostic of the details of the user's mental activity and does not rely on any data except timestamps of user's interactions, accounting for individual user differences. The method is implemented[1] in R and has been tested in the data from a large sample of HarvardX MOOCs.

## ACM Classification Keywords
I.5.m. Pattern Recognition: Miscellaneous; I.6.5. Simulation and modeling: Model development; K.3.1: Computers and education: Computer uses in education;

## INTRODUCTION
Recognizing when a learner wanders off task is often a difficult matter even in face-to-face learning, and in distant learning it is even more difficult. Focusing on estimating only the total amount of time-on-task is a simpler challenge and there is a clear hope that the method can be extended from it to the problem of detecting off-task behavior in real time. This opens up the possibility of interventions in online learning environments: users with low time-on-task are at risk of dropping out or failing the course and they can be contacted. Furthermore, the synchronous incremental time-on-task can be used as a variable for recommendation algorithms in adaptive learning.

The importance of the time-on-task as a learner's parameter has been noted a long time ago, first in the traditional off-line learning [1,2]. When learners interact with educational software, such as in a MOOC, the time-on-task needs to be inferred from the user's track log data. One can study the patterns in the timestamps of the user's actions, possibly coupled with the details of what those actions are. Very typically, the studies of the timestamps involve thresholds. For instance, the study [3] operates with several thresholds, from 60 to 180 seconds. Fundamentally, the idea of this approach is simple: a longer time interval between user's consecutive actions is an indication that the user went off-task. The question that we are trying to answer in this work is: can we avoid setting thresholds heuristically, either by adopting a method that does not need them, or by finding a data-drive way to set the threshold values?

---

[1] The code is open-source: https://github.com/harvard-vpal/time-on-task-GMM

Given the complexity of the learners' behaviors, we deliberately suggest a simple phenomenological approach for estimating the time-on-task: we base it on the statistics of the user's observed time signature of clicks. A secondary, pragmatic advantage of this approach is the common availability of track log data. Some educators may find it challenging to obtain many special user variables, but simple track log data is exported by most learning management systems.

## GENERAL METHOD
Consider the track log of user activity in the course: a chronologically ordered list of timestamps $t_i$ of user's clicks. Taking the differences between the consecutive timestamps we obtain a sequence of the user's time intervals $\Delta_i = t_i - t_{i-1}$, where $i = 1,2,\ldots N$. The core idea is to regard the set of user's time-intervals as a sample of a random variable, whose distribution is a mixture of two components, entirely-on-task and those that are at least partially off-task and tend to be longer (for brevity, we will refer to them as "off-task"). Let the probability densities of the two components be $\phi_k(t)$, with mean values $M_k$. Here $k = $ "on","off" and we assume that $M_{\text{on}} < M_{\text{off}}$. Fitting a mixture model to our data values $\Delta_i$ we will identify the distribution parameters and obtain a $N \times 2$ membership matrix $[p_{i,k}]$, each entry of which is the probability that a time-interval $\Delta_i$ belongs to the component $k$ ($\sum_k p_{i,k} = 1$ for any $i$). In particular, it means that $p_{i,\text{on}}$ is the vector of probabilities that $\Delta_i$ were spent entirely on-task.

The expectation value of the time-on-task is the product of the number of intervals $N$ and the mean value of the entirely-on-task distribution component, i.e. simply

$$T = NM_{\text{on}} \qquad (1)$$

In a little more detail, $M_{\text{on}}$ is the expectation value of the duration of an entirely-on-task interval. But we also assume that an off-task intervals contain on-task subintervals, and it is natural to estimate their average duration as also $M_{\text{on}}$. One might say that we extend the expected duration of "entirely-on-task" intervals to all intervals. Obviously, we can do that only as long as we assume that $M_{\text{on}} < M_{\text{off}}$: the expectation value of subinterval length must be smaller than that of full intervals.

In this approach, the mixture model is fitted for each user's data independently. It may prove too heavy computationally to do this in real time (e.g. for daily or weekly data reports on tens of thousands of MOOC users). A much faster method is to set a threshold $\tau$, and declare that an interval is entirely-on-task if the interval is shorter than $\tau$. In full analogy with Eq. 1, the total time-on-task is then estimated as

$$T'(\tau) = N \frac{\sum_{i=1}^{N} \Delta_i \mathbf{1}_{\Delta_i < \tau}}{\sum_{i=1}^{N} \mathbf{1}_{\Delta_i < \tau}} \qquad (2)$$

Thus, we may resort to this formula, but use the mixture model on historical data from similar courses as a way to determine the effective threshold $\tau$. Namely, we will find such $\tau$ that brings Eqs. 1 and 2 in agreement on historical data. Loosely speaking, the effective threshold is found as the zero of the function $F(\tau') = T'(\tau') - T$. One needs to be careful, however, because the function $F(\tau')$ is non-decreasing but also discontinuous: it consists of segments of constant value (shelves) separated by discontinuities. So, strictly speaking, it may not have a zero at all, but rather a value $\tau$ at which $F$ experiences a discontinuity jump from a negative to a positive value. A third, rather unlikely, possibility is that one of the shelves of $F$ has value exactly 0, in which case it is not important which point within the shelf we choose as $\tau$ and we can agree to take the midpoint. Furthermore, solving $F(\tau) = 0$ gives an individual effective threshold $\tau$ for each user. As a final step, these user-specific thresholds can be averaged to end up with a single threshold value[2].

Hence, the overall strategy for estimating the total time-on-task is two-fold. If computational time is not an issue, we fit a mixture model to each user's time intervals in the track log and use Eq. 1. Otherwise, we use Eq. 2, where the threshold value $\tau$ is found by fitting the mixture model, solving $T = T'(\tau)$ for every user in the historical data and averaging across the users. By "solving" we mean the procedure with the caveats described above.

**LOG-NORMAL MIXTURE MODEL**
The observed $\Delta_i$'s tend to have a very skewed distribution, qualitatively similar to a log-normal distribution. (Indeed, when the time intervals are spent on similar tasks (e.g. the response times for assessment questions), there is a reason to expect a log-normal distribution on the grounds of the central limit theorem.) We therefore fit them with a mixture of log-normal distributions. This choice is convenient: switching to the logarithms of time-intervals as our new variables, we need to fit to them the $K$-component Gaussian mixture model (GMM) – perhaps the most commonly used mixture model of all [5]. The probability density of the logarithm of a time interval is

$$P(\log \Delta) = \frac{1}{\sqrt{2\pi}} \sum_{k=1}^{K} \frac{a_k}{\sigma_k} \exp\left(-\frac{(\log \Delta - \mu_k)^2}{2\sigma_k^2}\right),$$

where $\sum_{k=1}^{K} a_k = 1$. It may seem that our method dictates using a 2-component GMM, but this is not so because we don't have to demand that the "entirely-on-task" and "off-task" components each consist of a single Gaussian component. Generally, if the data is fitted with a $K$-component GMM, we will classify them into two groups: the "entirely-on-task" group and the "off-task" group. The question of choosing $K$ remains but, given a value, the GMM produces a membership matrix $p_{i,k}$, and from that the means and standard deviations of the Gaussians, as well as their mixing coefficients: $\mu_k, \sigma_k, a_k$ for $k = 1, ..K$.

Returning from the logarithms to the time-intervals themselves, we are interested in the mean values of the log-normal components in the distribution of $\Delta_i$. If we were dealing with the parameters of true log-normal distributions, these mean values would be $m_k = \exp(\mu_k + \sigma_k^2/2)$. However, since we are dealing with sample distributions, in some cases the application of this formula causes a mismatch, such that Eq. 2 yields a nonsensical result greater than the user's net time $\sum_i \Delta_i$. A better strategy is to calculate the mean values directly:

$$m_k = \frac{\sum_{i=1}^{N} \Delta_i p_{i,k}}{\sum_{i=1}^{N} p_{i,k}}$$

Let us adopt the convention that the GMM components are indexed in the order of increasing $m_k$.

A reasonable model is $K = 3$, reflecting three different user behaviors: very short time-intervals result from thoughtless clicking-through or guessing (in case of assessment items), the medium time-intervals result from "thoughtful interactions", and the very long time-intervals result from the user getting off-task (distracted, or even leaving). We may classify the Gaussian components 1 and 2 as "entirely-on-task" and 3 as "off-task". More generally, we can determine $K$ using the Bayesian information criterion, and declare all the components except the last one ($k = 1, ..K - 1$) "entirely-on-task". Then Eq. 1 for time-on-task becomes[3]

$$T = N \frac{\sum_{k=1}^{K-1} a_k m_k}{\sum_{k=1}^{K-1} a_k} \qquad (3)$$

As a goodness-of-fit measure of the GMM for each user, we compare the observed cumulative distribution function $\Phi(x)$ of logarithms of time-intervals and the fit distribution function $\Phi_{GMM}(x)$, taking the values of both functions with the observed $\log \Delta_i$ as arguments. We use the Pearson correlation (R-squared) as the goodness-of-fit measure:

$$\text{g.o.f.} = \rho\big(\Phi(\log \Delta_i), \Phi_{GMM}(\log \Delta_i)\big) \quad (4)$$

---

[2] This last step is amenable to many modifications. A simple average could be replaced by many alternatives. Also, one could break users into cohorts based on some engagement parameter, such us test grades, and produce different thresholds for each cohort. Or else, a regression $\tau(X)$ with respect to any number of user variables $X$.

[3] Including fast clicks of component 1 into the on-task behavior is our choice. If, however, we wished to remove the component 1 from the time-on-task, the formula is modified to $T = (N - \sum_{i=1}^{N} p_{i,1}) \sum_{k=2}^{K-1} a_k m_k$.

**AN APPLICATION OF THE MODEL**

We created a function in R to perform the analysis of course track logs, which will be made open source. At its core is fitting a GMM model by an iterative Expectation-Maximization algorithm in R. For a very few users (a fraction of a percent) the algorithm did not converge and we drop those users from the subsequent analysis.

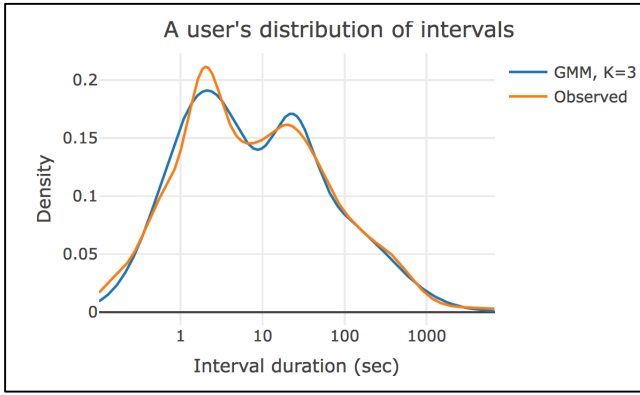

**Figure 1. An example of a GMM fit for one user. The observed density is obtained from the data with a Gaussian kernel.**

Click timestamps from 52 HarvardX courses from the year 2016 were analyzed. Users with less than 20 clicks were discarded from the data, as were the intervals $\Delta_i$ shorter than 0.1 seconds or longer than 2 hours. For each user, we attempted a fit with 3, 4, or 5 Gaussian components, choosing the optimal fit by minimizing the Bayesian information criterion [6]. Most of the time the criterion chose 3 components. For instance, in one course the breakdown of fitting with 3/4/5 components was 83%/15.3%/1.6% of users.

We might expect that the statistics of intervals between clicks should depend on the resource type (perhaps, people click less frequently in videos than in problems, etc.). To address this concern, we also fitted each user's intervals separately by the type of resource (e.g. a video, or an assessment question), to check how much it affects the results. A time interval is categorized as happening on a particular resource type, e.g. videos, if both clicks occurred on such resource; time intervals with clicks occurring on different resource types are categorized as mixed, e.g. video/question time intervals.

We take the mean of Eq. 4 across users in every course, subtract from it one standard deviation to err on the side of caution, and report the result as an aggregate goodness-of-fit measure for the course as a whole. In all courses (for any resource type separately or for all together), this aggregate measure was 0.994 or higher, indicating a good agreement of the model with the data.

In Figure 2 we show the effective thresholds $\tau$ calculated for each course. Moreover, in each course we calculate it from several different fits of GMM: all intervals together, as well as separately per resource type. The fact that the value changes little because of that is the reason we believe that our mixture modeling captures different mental states of the learner, rather than simply reflect the differences in clicking patterns between resource types.

Going forward, these findings will enable us to use the quick method of Eq. 2 to track time-on-task in any HarvardX course: we have determined the value of $\tau$ from the 2016 historical data. We could aggregate it by course topic, but there is no strong dependence on it in Figure 2, so it makes sense to simply average the values across all 52 courses, which comes out to be $\tau = 322$ seconds. At the same time, Figure 2 shows a substantial variability in effective thresholds across courses, indicating that using Eq. 2 with one threshold for any course and user is, like most "one size fits all" approaches, suboptimal.

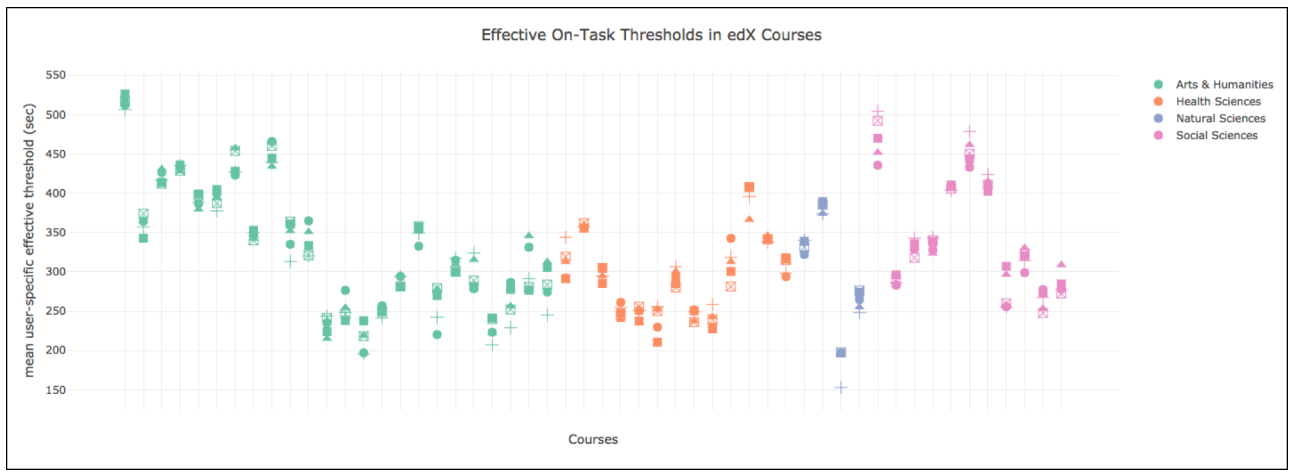

**Figure 2: User-averaged effective on-task threshold $\tau$. The courses are color-coded by their topic. For each course, different symbols mark the threshold as obtained from different subsets of intervals. Filled squares – problems; squares with crosses – videos, crosses – sequentials; triangles – other resource types; circles – all resource types indiscriminately.**

In addition to the effective threshold, our R routine outputs for each user the total estimated time-on-task $T$, the ratio of it to the net time, the average duration of an "entirely-on-task" interval, and many other, more technical, quantities. For instance, averaged across the courses, the fraction of net time spent by a user on task turns out to be 19%.

Relation of time-on-task to other markers of performance in distance learning is not the subject of this research. But since the output of our method is the total time-on-task calculated for each user, it is easy to join it with other user variables and explore the connections. As a first non-trivial application of this sort, we find the Pearson correlation between the logarithm of the time-on-task and the final grade in the course. Averaged across our courses, this correlation value turned out to be 0.49, which is quite encouraging. The study [4] shows, in a setting quite different from edX, the $R^2$ values for the final grade as no more than 0.28. It is not a matter of straightforward comparison of the results (for one thing, it is far from clear that a better estimate of the time-on-task must have a higher correlation with the grade). Rather, we quote this study's result to show what values might be reasonably expected. Moreover, by averaging across courses the group averages of logarithms of time-on-task, we find that those users who completed a course, had an average of 7.5 times longer time-on-task compared to those who did not.

## CONCLUSIONS

We have created an open-source R routine for estimating the time-on-task of users in a learning platform. As the input, it uses the track log of user clicks: a data table with a column of usernames and a column of timestamps. The estimation of time-on-task is done for each user individually using a Gaussian mixture model, without imposing external thresholds, in a data-driven way. We applied the method to the historical track logs from a large number of HarvardX courses.

Given the computational complexity of the model (in our experience, processing one course took between several hours and a day, depending on the course size), it may be necessary to resort to a simplified calculation with a time threshold for the inter-click intervals if we are interested in estimating time-on-task in real time or thereabouts. In this case, applying our mixture model to the historical data of similar courses provides the threshold value (in our group of HarvardX courses we found it to be 322 seconds). Moreover, it is easy to specialize the threshold calculation as needed, for example, to use different thresholds for courses on different topics, or for different cohorts of users. Since our method calculates the effective threshold individually for each user in the historical data, it is a matter of aggregating the results by cohort and by course topic.

The model is validated by the high goodness of fit. Furthermore, as expected, the time-on-task it produces is strongly and positively correlated with completing the course and with the final grade. We see the separation of components in the mixture of time intervals clearly. We also see that it is not simply due to the interaction with different resource types. Therefore, it stands to reason that this separation is a reflection of the differences in a user's several different modes of operating. But being purposely phenomenological, this model leaves one important point unaddressed: how do we know that these modes are specifically the on-task behavior versus the off-task behavior? This question is inseparable from the fundamental difficulty of the entire time-on-task discussion: lack of a universal and operational definition of on-task and off-task behavior in different online settings. Combining our method with direct observation of learners (eye-tracking techniques, etc. [7]) might provide some insights in the future.

## ACKNOWLEDGEMENTS
I am grateful for the support from the Office of the Vice Provost for Advances in Learning at Harvard University. Many thanks to Glenn Lopez for his help with course data.

## REFERENCES

1. Karweit, N. and Slavin, R.E., 1982. Time-on-task: Issues of timing, sampling, and definition. *Journal of educational psychology*, *74*(6), p.844.

2. Stallings, J., 1980. Allocated academic learning time revisited, or beyond time-on-task. *Educational researcher*, *9*(11), pp.11-16.

3. Grabe, M. and Sigler, E., 2002. Studying online: evaluation of an online study environment. *Computers & Education*, *38*(4), pp.375-383.

4. Kovanović, V., Gašević, D., Dawson, S., Joksimović, S., Baker, R.S. and Hatala, M., 2015, March. Penetrating the black box of time-on-task estimation. In *Proceedings of the Fifth International Conference on Learning Analytics And Knowledge* (pp. 184-193). ACM.

5. Lindsay, B.G., 1995, January. Mixture models: theory, geometry and applications. In *NSF-CBMS regional conference series in probability and statistics* (pp. i-163). Institute of Mathematical Statistics and the American Statistical Association.

6. Schwarz, G., 1978. Estimating the dimension of a model. *The annals of statistics*, *6*(2), pp.461-464.

7. Zhao, Y., Lofi, C. and Hauff, C., 2017, September. Scalable Mind-Wandering Detection for MOOCs: A Webcam-Based Approach. In *European Conference on Technology Enhanced Learning* (pp. 330-344). Springer, Cham.